\documentclass{appolb}
\usepackage{amsmath}
\usepackage{amssymb}
\usepackage{array}
\usepackage{calc}
\usepackage{longtable}
\usepackage{multirow,booktabs}
\usepackage{relsize}
\usepackage{pstricks}
\usepackage{graphicx}
\usepackage{xspace}
\usepackage{units}
\usepackage{hyperref}
\newcommand{\MCatNLO}{M\protect\scalebox{0.8}{C}@N\protect\scalebox{0.8}{LO}\xspace}

\newcommand{\MEPSatLO}{M\protect\scalebox{0.8}{E}P\protect\scalebox{0.8}{S}@L\protect\scalebox{0.8}{O}\xspace}
\newcommand{\MEPSatNLO}{M\protect\scalebox{0.8}{E}P\protect\scalebox{0.8}{S}@N\protect\scalebox{0.8}{LO}\xspace}

\newcommand{\Sherpa}{S\protect\scalebox{0.8}{HERPA}\xspace}

\newcommand\ATLAS{\atlas}
\newcommand\atlas{A\protect\scalebox{0.8}{TLAS}\xspace}
\newcommand\LHC{L\protect\scalebox{0.8}{HC}\xspace}

\begin{document}
\title{$b$ quark mass effects in associated production%
\thanks{Presented at the HiggsTools Final Meeting, Durham}%
}

\author{Davide Napoletano$^{1,2}$\\
  {\tiny $^1$IPhT, CEA Saclay, CNRS UMR 3681, F-91191, Gif-Sur-Yvette, France}\\
  {\tiny $^2$Institute for Particle Physics Phenomenology,
  Durham University, Durham DH1 3LE, UK}\\
  \vspace{0.5cm}}
\maketitle
\begin{abstract}
  In this work we study an extension of the commonly used 5F scheme, where $b$ quarks
  are treated as massless partons, in which full mass effects are retained in both
  the initial and in the final state. We name this scheme 5F massive scheme (5FMS).
  We implement this scheme in the \Sherpa Monte Carlo event generator
  at \MEPSatNLO accuracy, and we compare it for two relevant cases for the \LHC:
  $b\bar{b} \rightarrow H$ and $pp\rightarrow Zb$.
\end{abstract}

\PACS{PACS numbers come here}
  
\section{Introduction}
Processes with heavy quarks in the initial state, in particular
associated production processes,
have seen in recent years a renewed interest~
\cite{Krauss:2016orf,Napoletano:2017czh,Forte:2015hba, Forte:2016sja,Maltoni:2012pa,Lim:2016wjo,Bonvini:2015pxa,
  Bonvini:2016fgf,Bertone:2017djs}.
From the theoretical point of view, they are interesting applications
of multiscale processes with largely different scales.
Ratio of these large scales, can give rise to large logarithms
which might spoil the convergence of the perturbative series.
To avoid this, one can consider the $b$ as a massless parton,
and construct a $b$-PDF which resums this potentially large
collinear logarithms, at the price of neglecting
mass effects.
An alternative point of view can be that of treating 
the $b$-quark as a massive, decoupled particle,
which is only produced in the final state, or treating the $b$-quark
as a massless parton on the same footing as the other, thus contributing
to the QCD evolution. In this way one is able to retain full mass effects
at the price of keeping the aforementioned possibly large collinear logs.

\begin{figure}[htb]
  \centering{%
    \raisebox{-0.6cm}{\includegraphics[scale=0.2]{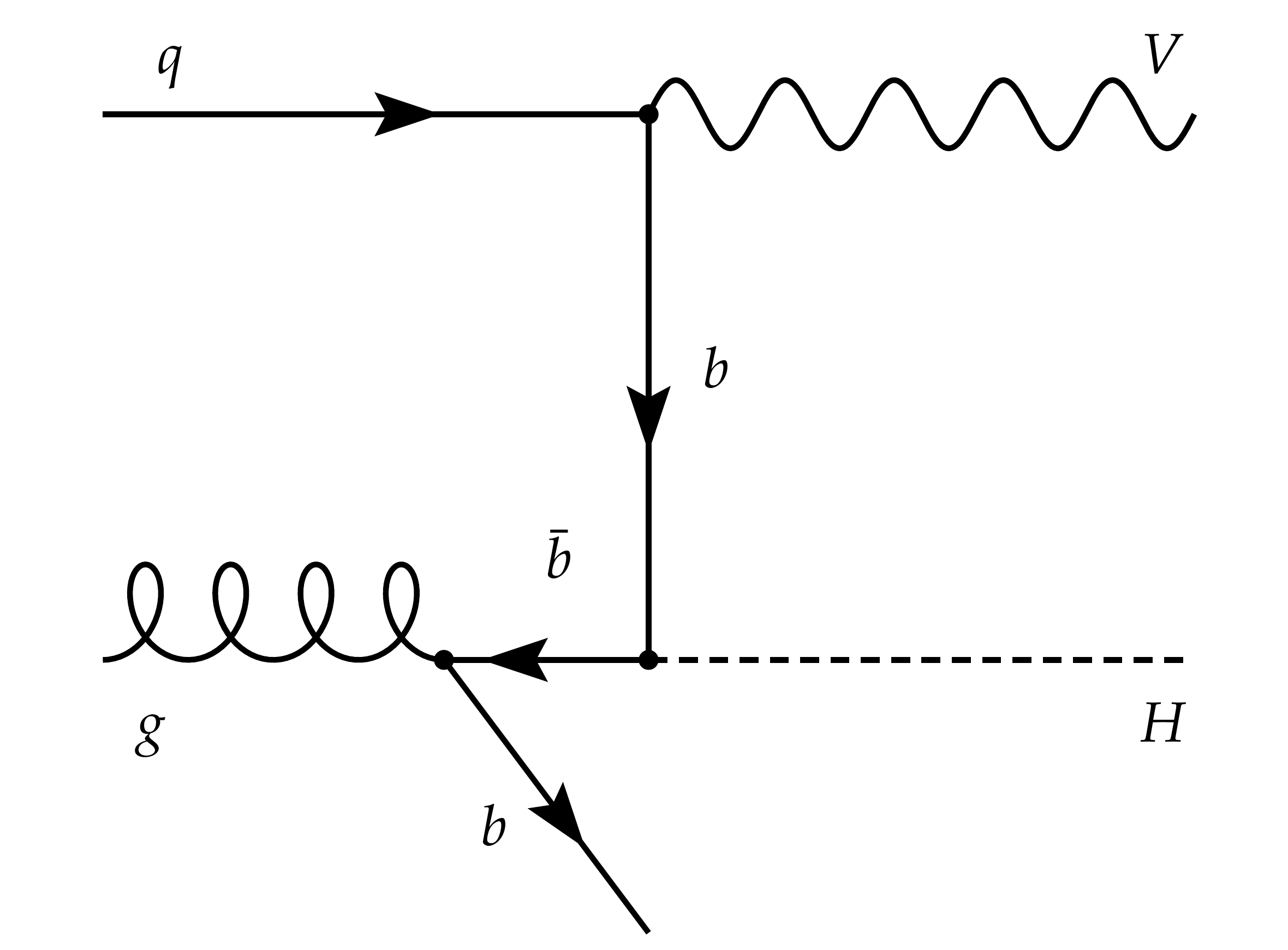}\hspace{2.5cm}}
    \includegraphics[scale=0.2]{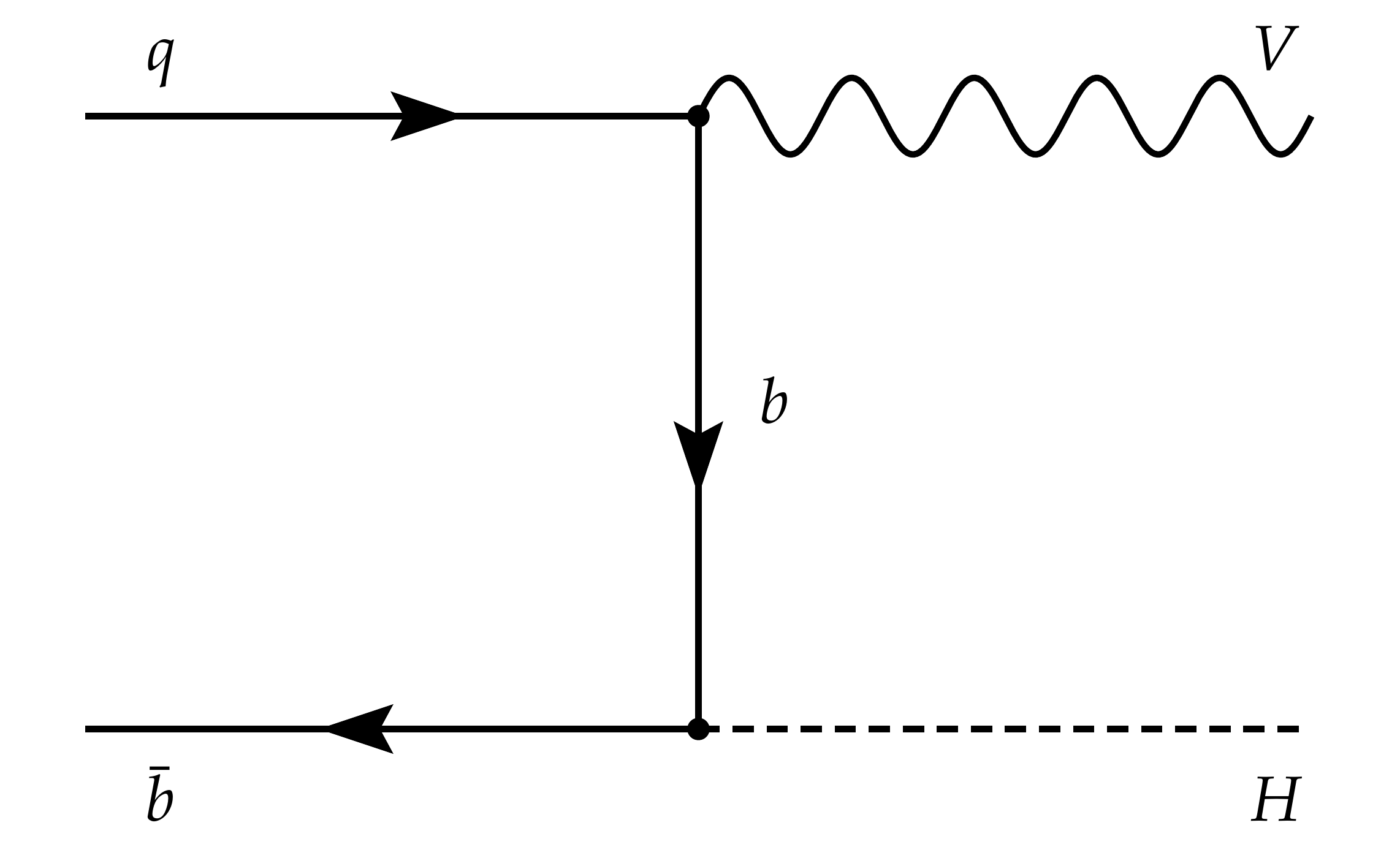}}
  \caption{4F (left plot) vs 5F (right plot) scheme diagrams for $VH$ production}
  \label{Fig:4Fvs5F}
\end{figure}
\begin{figure}[htb]
  \begin{center}%
    \begin{tabular}{cc}
      \includegraphics[width=0.5\textwidth]{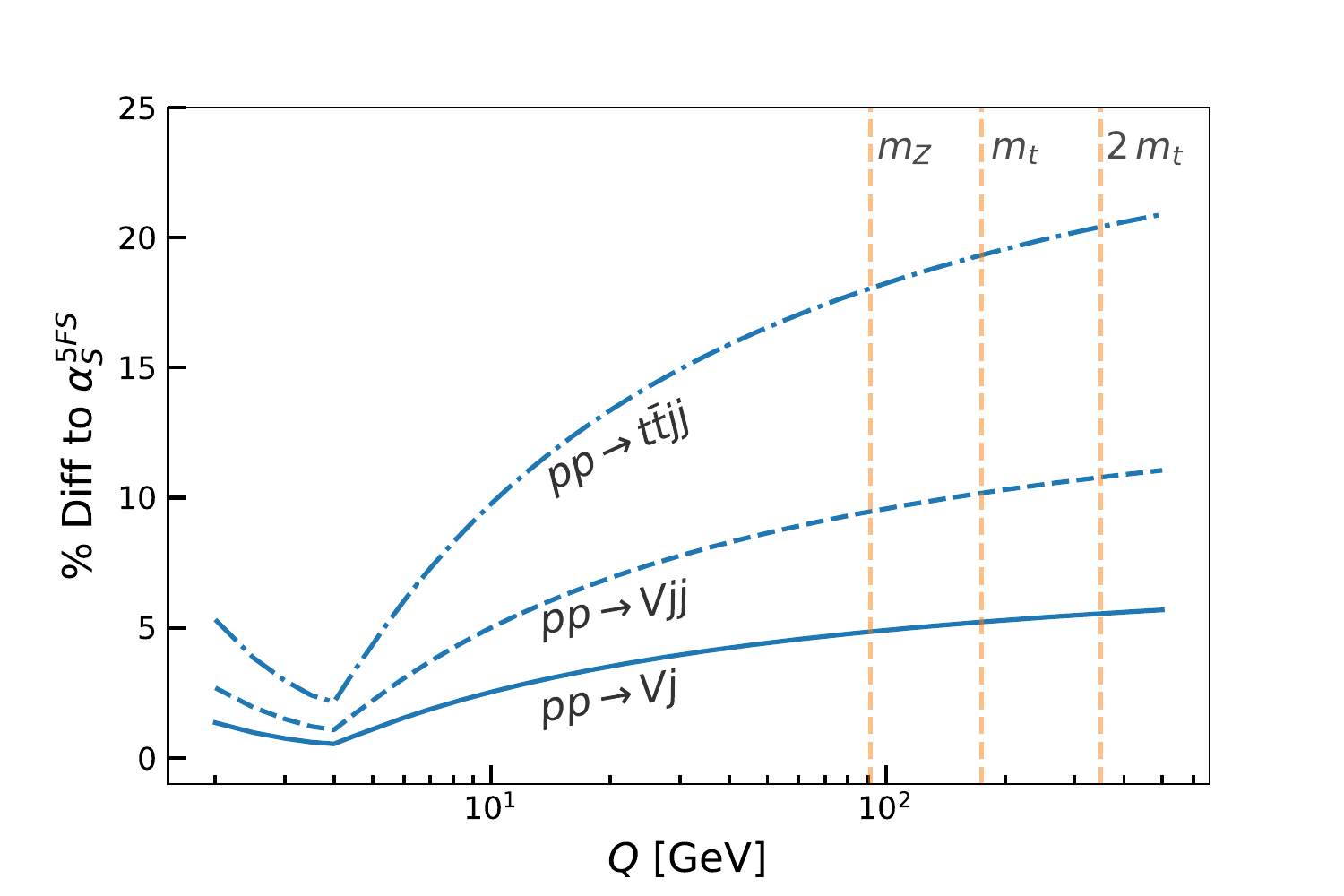} &
      \includegraphics[width=0.5\textwidth]{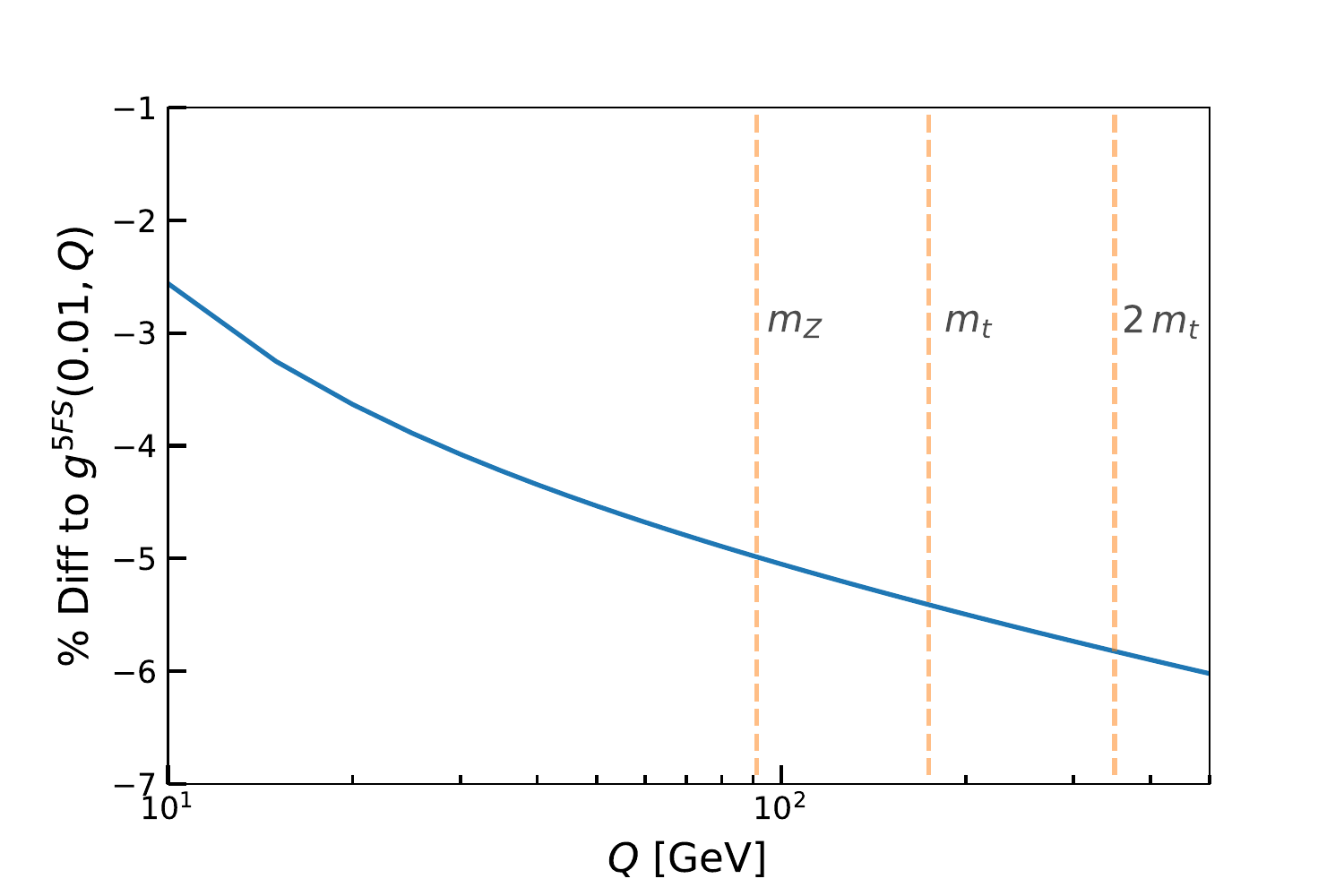}
    \end{tabular}
    \caption{In the plot is shown the error that is made when taking and $\alpha_s$
      and a gluon PDF in the 4FS with respect to the 5FS baseline. As it can be seen
      the two effects partially mitigate each other, although this is true only
      for processes that start at a low enough power of $\alpha_S$, and have a large
      gluon contribution.}
    \label{Fig:alphapdf}
  \end{center}
\end{figure}

The former of these two approaches is called {\it five-flavour} (5F) scheme
and would schematically corresponds to the right hand side plot of~Fig.\ref{Fig:4Fvs5F},
while the latter is refered to as {\it four-flavour} (4F) scheme and
is represented in the left plot of~Fig.\ref{Fig:4Fvs5F}.
These two approaches have generally been used in a complementary, with the
old way of saying being:
\begin{quote}
  `` {\it use the 4FS for exclusive observables, \\ and the 5FS for inclusive
    observables} "
\end{quote}

Many studies have however now shown that the 5FS scheme
performs generally better both when compared to data, \cite{Krauss:2016orf},
or when comparing it with a matched calculation \cite{Forte:2016sja,Bonvini:2016fgf},
although this too is only true up to a certain extent. There are,
in fact regions of phase space where one might still want to include
exact mass effects, which would in principle require the use of the 4FS.

In this work we investigate the possibility of using a scheme, built upon the 5FS,
with exact mass dependence. We name this scheme five-flavour-massive-scheme (5FMS).
We implement the necessary ingredients to perform calculations in
this scheme in the \Sherpa Monte Carlo event generator~\cite{Gleisberg:2008ta},
at \MCatNLO accuracy~\cite{Gehrmann:2012yg,Hoeche:2012yf}.
A detailed description of this scheme and its implementation
can be found in~\cite{Krauss:2017wmx}.

\section{Including mass effects}
\subsection{Fixed order}
\begin{figure}[!htb]
  \centering{%
    \includegraphics[width=12.5cm]{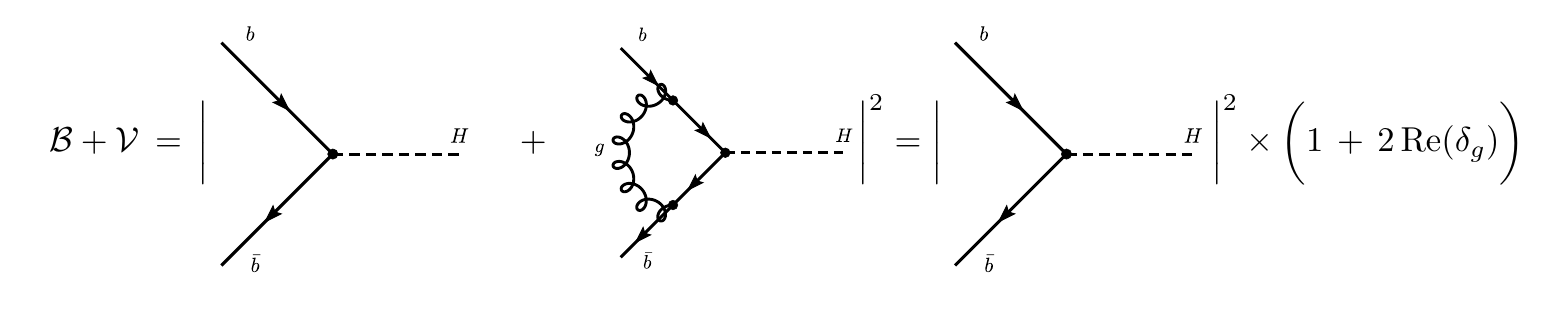}
    \includegraphics[width=12.5cm]{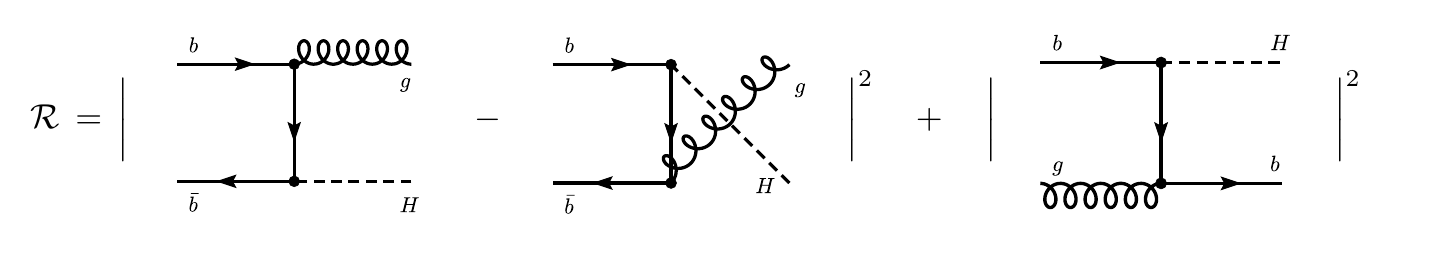}}
  \caption{Virtual and Real contriutions to $b\bar{b}\rightarrow H$}
  \label{Fig:bvr}
\end{figure}

\begin{figure}[htb]
  \centering{%
    \includegraphics[width=0.6\textwidth]{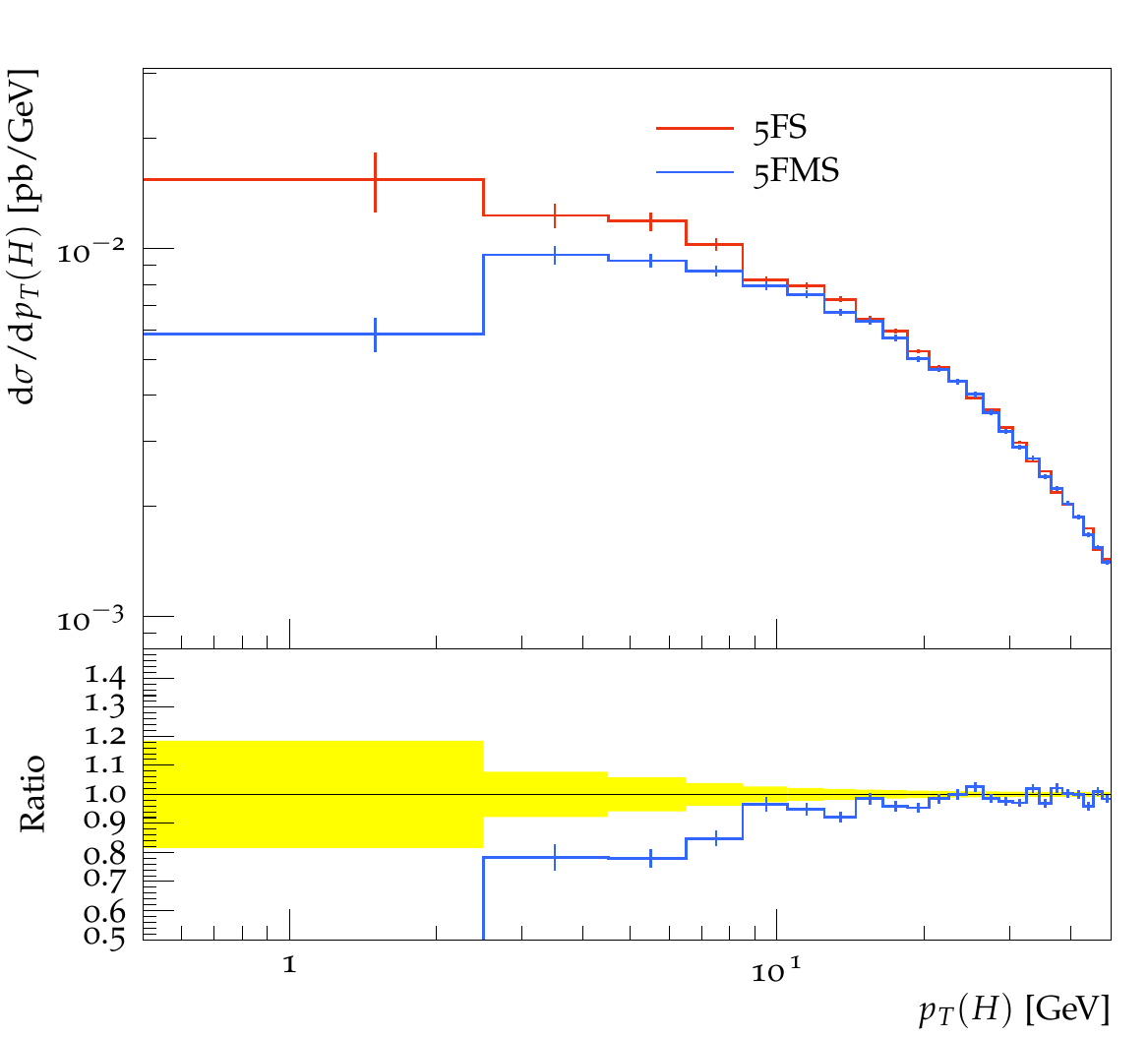}}
  \caption{Comparison of the 5F and the 5FM scheme.}
  \label{Fig:5F5FMS_fo}
\end{figure}

In order to study the effects introduced by this new scheme, we take an explicit example:
$b\bar{b} \rightarrow H$. Reference diagrams that contribute to the {\it next-to-leading} order
are shown in~Fig.\ref{Fig:bvr}. At the level of partonic matrix elements, the only
difference between the 5FS and the 5FMS is that in the latter full mass dependence
is retained, including in the initial state. As the infrared divergent structure is
modified by the presence of the $b$ mass, that acts as a collinear regulator,
a modification of the standard Catani-Seymour subtraction is required~\cite{Krauss:2017wmx}.
With this in place, we can generate {\it fixed-order} events, Fig.\ref{Fig:5F5FMS_fo}.
As an example observable, we focus on the $p_T$ of the produced $H$ boson.

We know that mass effects contribute only a few percent to the total
cross section for this process. In addition, we know that they
are power suppressed and we expect them to scale like $m_b^2/p_T^2$.
This is, indeed, roughly the behaviour shown in~Fig.\ref{Fig:5F5FMS_fo}.

\subsection{\MCatNLO}
\begin{figure}[!htb]
  \begin{center}
    \begin{tabular}{cc}
      \includegraphics[width=0.4\textwidth]{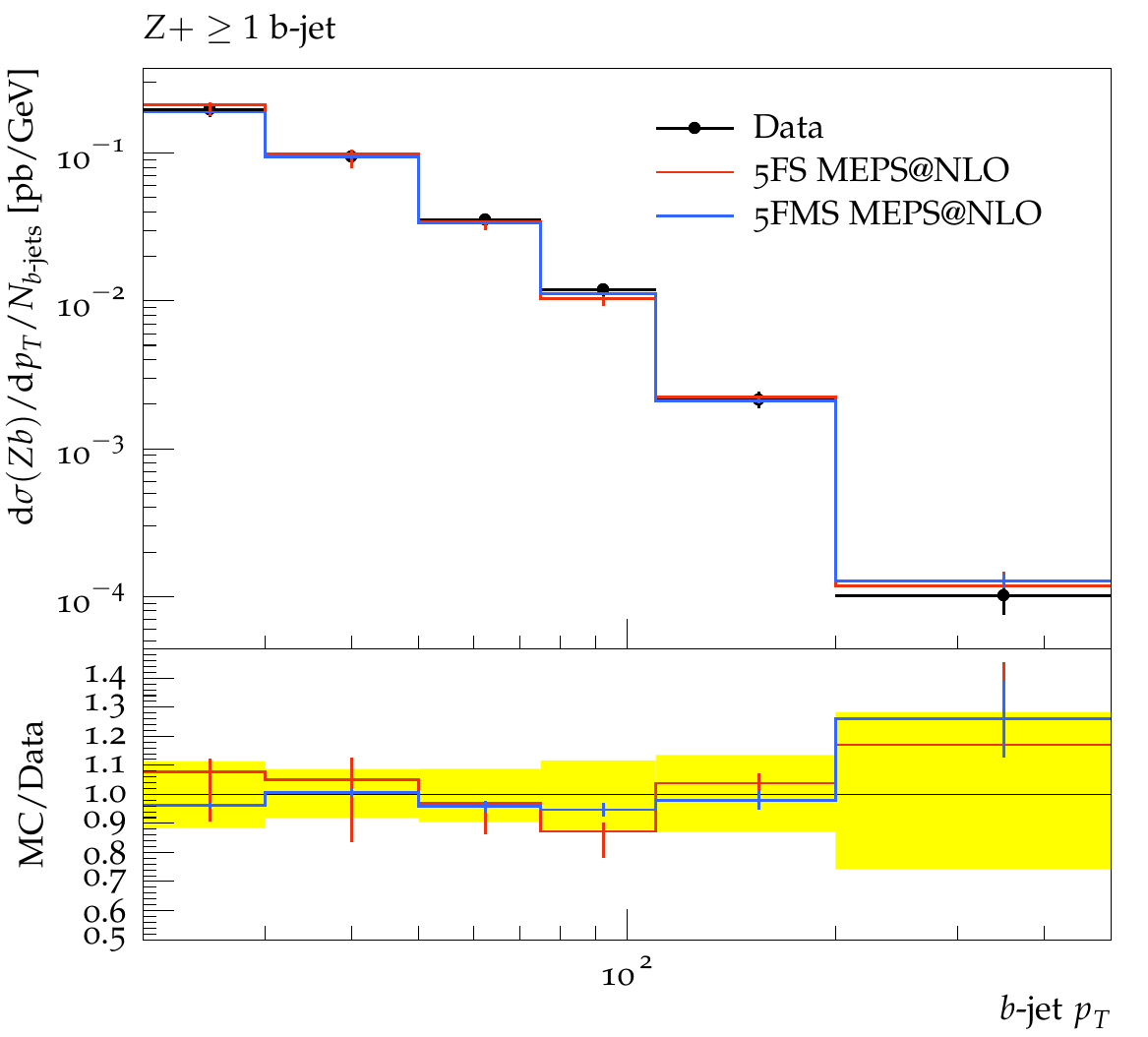}&
      \includegraphics[width=0.4\textwidth]{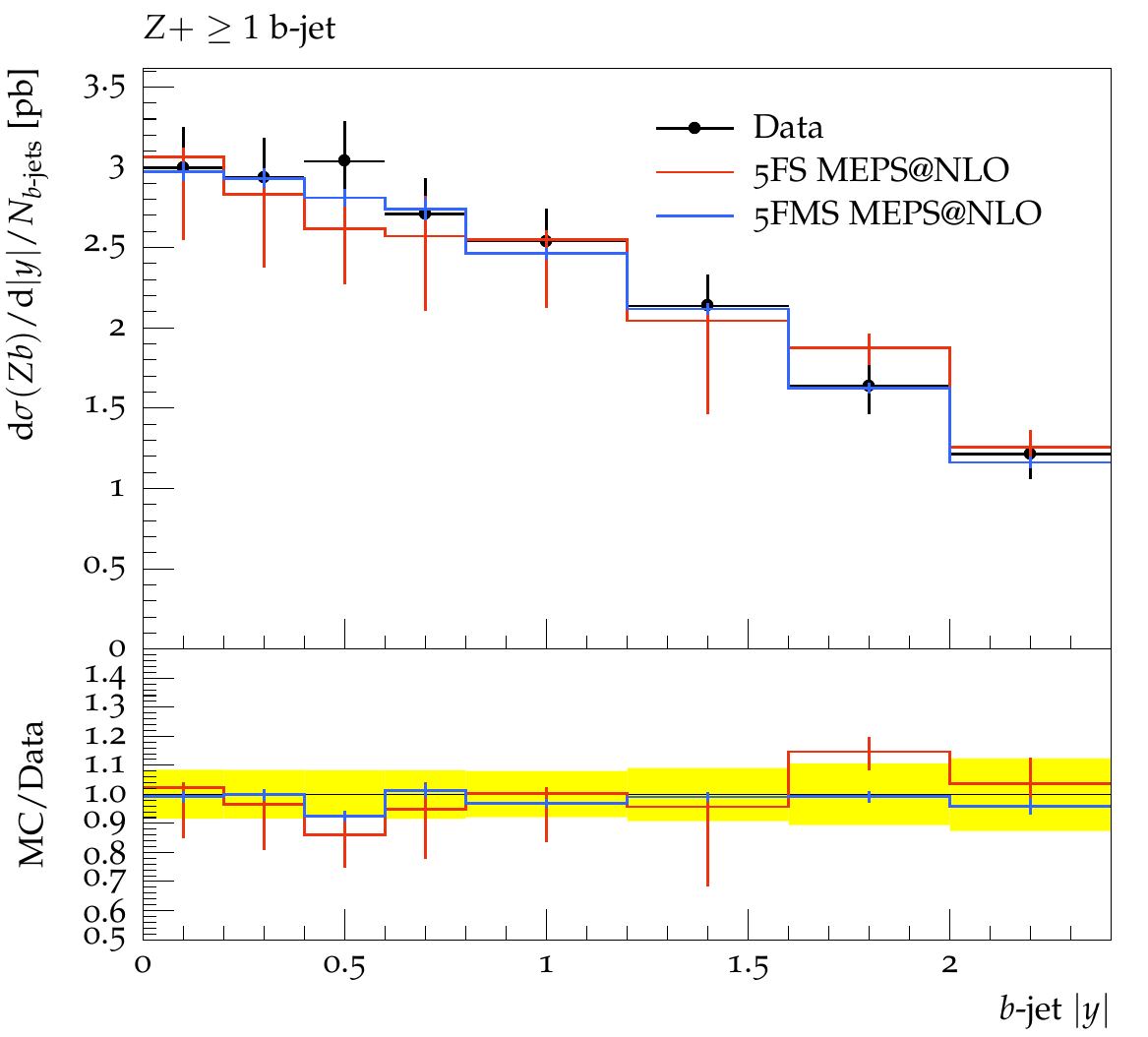}\\
      \includegraphics[width=0.4\textwidth]{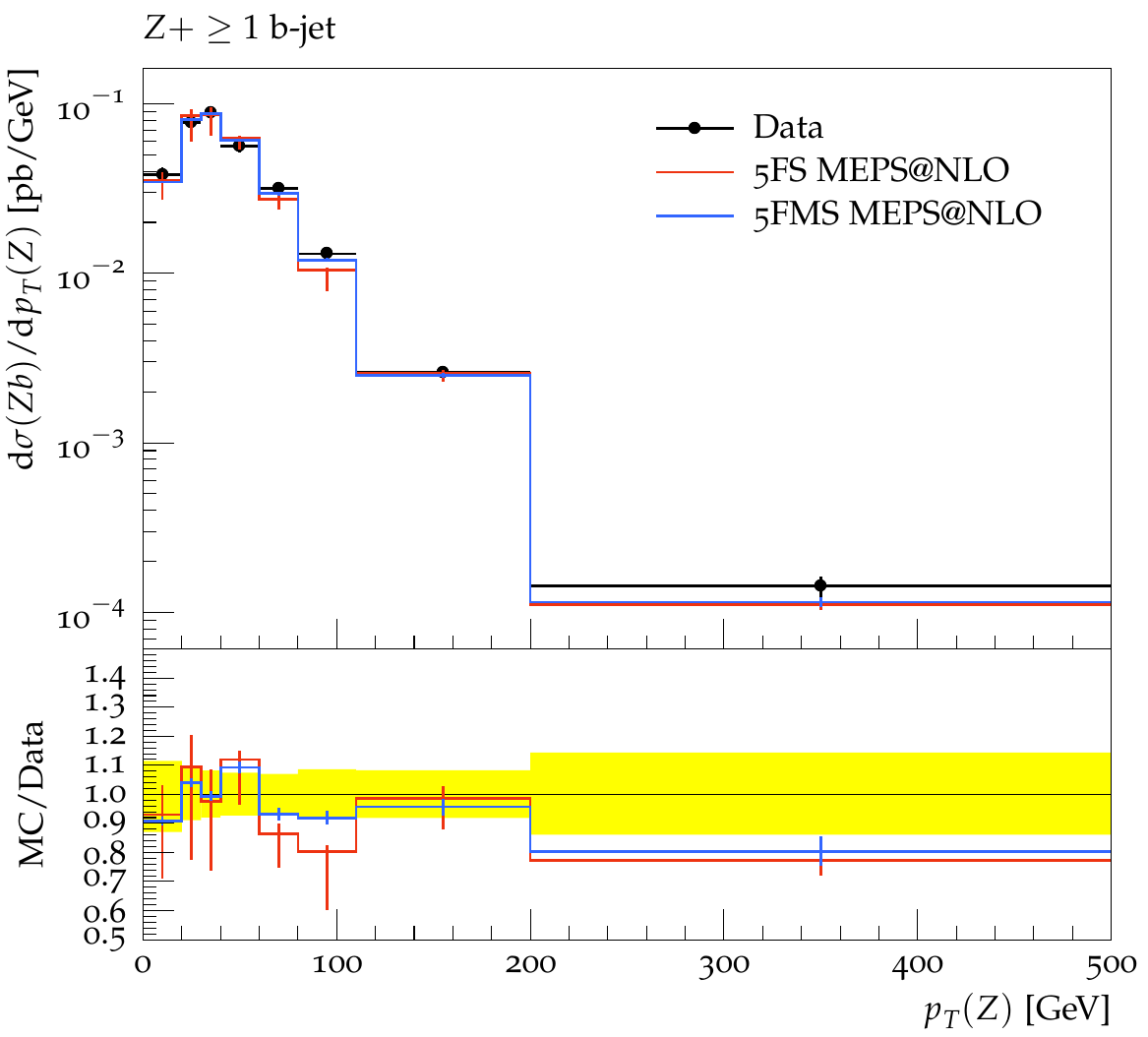}&
      \includegraphics[width=0.4\textwidth]{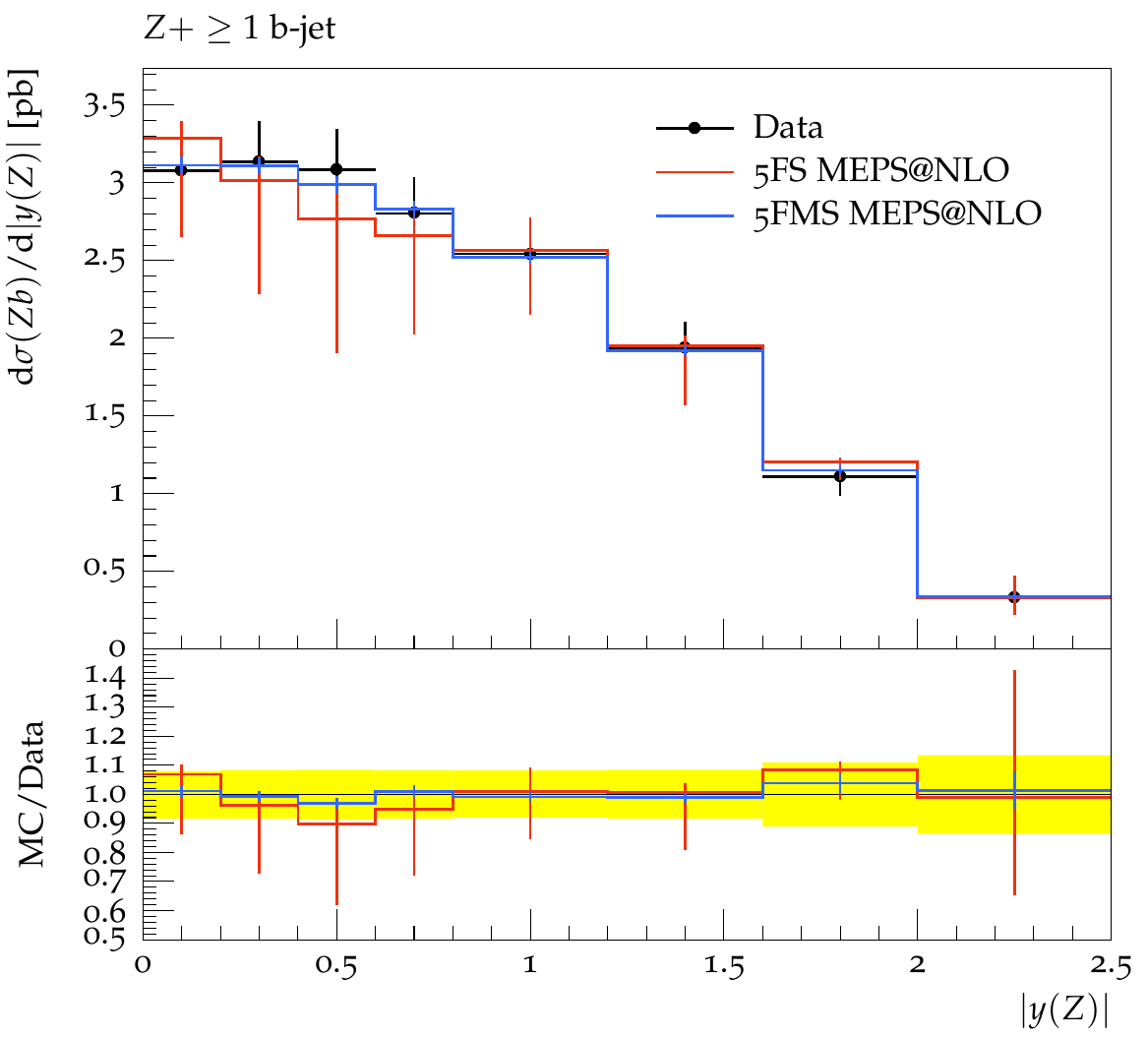}     
    \end{tabular}
    \caption{We show prediction obtained
      in the 5FS, massless, at \MEPSatNLO accuracy, with up to 2 jets at NLO
      plus up to three jets at leading order. The 5FMS prediction on the other hand
      includes only the 0 jet contribution at NLO, while the 1,2 and 3 jets
      contributions are merged with LO accuracy.}
  \label{fig:mcatnlo}
  \end{center}
\end{figure}

We now want to study what happens when this scheme is matched
to the parton shower. Since we don't have a theoretical reference here,
we use $pp\rightarrow Z b$ data~\cite{Aaboud:2017xsd} from \ATLAS.
In particular we replicate
the set-up used in~\cite{Krauss:2016orf}, and we compare with the
5FS \MEPSatNLO line referenced therein, see~Fig.\ref{fig:mcatnlo}.
The difference with respect to that set-up is that we have \MCatNLO
accuracy only for the core $pp\rightarrow Z$ processes, while
extra jet contributions that are merged on top of that only
come at leading order accuracy. Strictly speaking thus, we should
compare the 5FMS \MEPSatNLO here with the 5F \MEPSatLO prediction
of~\cite{Krauss:2016orf}, however we expect some mass effects to
make up for some of the differences in accuracy.

As our aim is to investigate mass effects, in $b$-initiated processes,
we look at events in which at least
one jet containing a $b$ is tagged, and we plot distributions
for the leading $b$-jet and the $Z$ boson $p_T$ and $y$ against data.
These plots are reported in~Fig.\ref{fig:mcatnlo}.
As it can be seen, this new scheme performs rather well, and, indeed,
it shows the same type of compatibility with data of the 5FS \MEPSatNLO
prediction, which is reassuring.

Further details and studies on this new scheme can be found in~\cite{Krauss:2017wmx}

\section*{Acknowledgements}
We want to thank our colleagues from the \Sherpa collaboration 
for fruitful discussions and technical support.
We acknowledge financial support from the EU research networks funded by the Research
Executive  Agency (REA) of the European Union under Grant Agreements 
PITN-GA2012-316704 (``HiggsTools'') and PITN-GA-2012-315877 (``MCnetITN''),
by the ERC Advanced Grant MC@NNLO (340983), and from BMBF under contracts
05H12MG5 and 05H15MGCAA,

\end{document}